\documentclass[a4paper,fleqn]{cas-dc}

\usepackage[numbers]{natbib}
\usepackage{slashed}
\usepackage{subfigure}

\def\tsc#1{\csdef{#1}{\textsc{\lowercase{#1}}\xspace}}
\tsc{WGM}
\tsc{QE}
\tsc{EP}
\tsc{PMS}
\tsc{BEC}
\tsc{DE}

\begin{document}
\let\WriteBookmarks\relax
\def\floatpagepagefraction{1}
\def\textpagefraction{.001}
\shorttitle{Hybrid renormalization in lattice calculation of baryon LCDAs}
\shortauthors{M.-H. Zhang}

\title [mode = title]{Hybrid renormalization in lattice calculation of baryon LCDAs}                      
\tnotemark[1]

\tnotetext[1]{Proceeding of the 2025 International Conference on the Structure of Baryons (Baryons 2025), 10-14 November 2025, ICC, Jeju, South Korea}

\author[1]{Mu-Hua Zhang}[type=editor,
                        orcid=0009-0005-8618-7172]
\fnmark[1]
\ead{muhuazhang@sjtu.edu.cn}


\affiliation[1]{organization={Tsung-Dao Lee Institute \& State Key Laboratory of Dark Matter Physics, Key Laboratory for Particle Astrophysics and Cosmology (MOE), Shanghai Key Laboratory for Particle Physics and Cosmology, School of Physics and Astronomy, Shanghai Jiao Tong University},
                addressline={800 Dongchuan RD.}, 
                city={Shanghai},
                postcode={200240},
                country={China}}

\fntext[fn1]{Speaker}

\begin{abstract}
At the 2025 International Conference on the Structure of Baryons (Baryons 2025), I presented our recent progress in lattice calculations of baryon light-cone distribution amplitudes (LCDAs). In Ref.~\cite{LatticePartonCollaborationLPC:2025vhd}, we implemented a novel hybrid renormalization scheme for octet baryons, leading to reliable determinations of quasi-distribution amplitudes (quasi-DAs).

The calculations were performed on $N_f=2+1$ ensembles with stout-smeared clover fermions and a Symanzik-improved gauge action at three lattice spacings, $a = 0.052,0.077,0.105$ fm. The hybrid renormalization removes linear divergences in lattice matrix elements and yields smooth, self-consistent quasi-DAs, providing a solid foundation for LaMET-based extractions of baryon LCDAs. Results at the continuum limit and physical pion mass will be reported in the near future.

\end{abstract}



\begin{keywords}
Lattice QCD \sep LCDA \sep LaMET \sep Baryon \sep Renormalization
\end{keywords}

\maketitle

\section{Introduction}

Light-cone distribution amplitudes (LCDAs) characterize the longitudinal momentum structure of partons inside hadrons and play a central role in the description of exclusive processes. The recent observation of CP violation in a baryonic system~\cite{LHCb:2025ray} has further highlighted the need for quantitatively reliable theoretical inputs, in particular for baryon LCDAs.

Early studies on baryon LCDAs date back to the asymptotic forms proposed in Ref.~\cite{Chernyak:1983ej} and the Chernyak--Ogloblin--Zhitnitsky model~\cite{Chernyak:1987nu}. Subsequent theoretical developments established the QCD framework for baryon LCDAs~\cite{Braun:1999te,Braun:2000kw} and studied baryon-octet LCDAs as well as flavor-symmetry breaking effects~\cite{Wein:2015oqa}.  Nevertheless, progress on first-principles calculations of baryon LCDAs has been comparatively limited. A major obstacle is that their definition involves light-cone correlations, which cannot be directly evaluated in Euclidean lattice QCD. Existing lattice efforts based on the operator product expansion (OPE) are restricted to the lowest moments~\cite{Bali:2015ykx,RQCD:2019hps,Bali:2024oxg}, and thus do not provide sufficient information for phenomenological applications.

The Large-Momentum Effective Theory (LaMET)~\cite{Ji:2013dva,Ji:2014gla} offers a practical approach to access the full momentum dependence of parton distributions from lattice QCD by relating equal-time correlators to light-cone observables. This framework has been successfully applied to meson LCDAs in recent years~\cite{Hua:2020gnw,LatticeParton:2022zqc,LatticeParton:2024zko}.

In this work, we report recent progress in the lattice determination of light-baryon LCDAs~\cite{LatticeParton:2024vck,LatticePartonCollaborationLPC:2025vhd}. Particular emphasis is placed on a hybrid renormalization strategy that removes linear divergences in lattice matrix elements and enables a controlled connection to the $\overline{\rm MS}$ scheme.

\section{LCDAs and Quasi-DAs for light baryons} \label{sec:framework}

\begin{figure*}
\centering
\subfigure[LCDAs in coordinate space.]{
    \centering
    \includegraphics[scale=0.12]{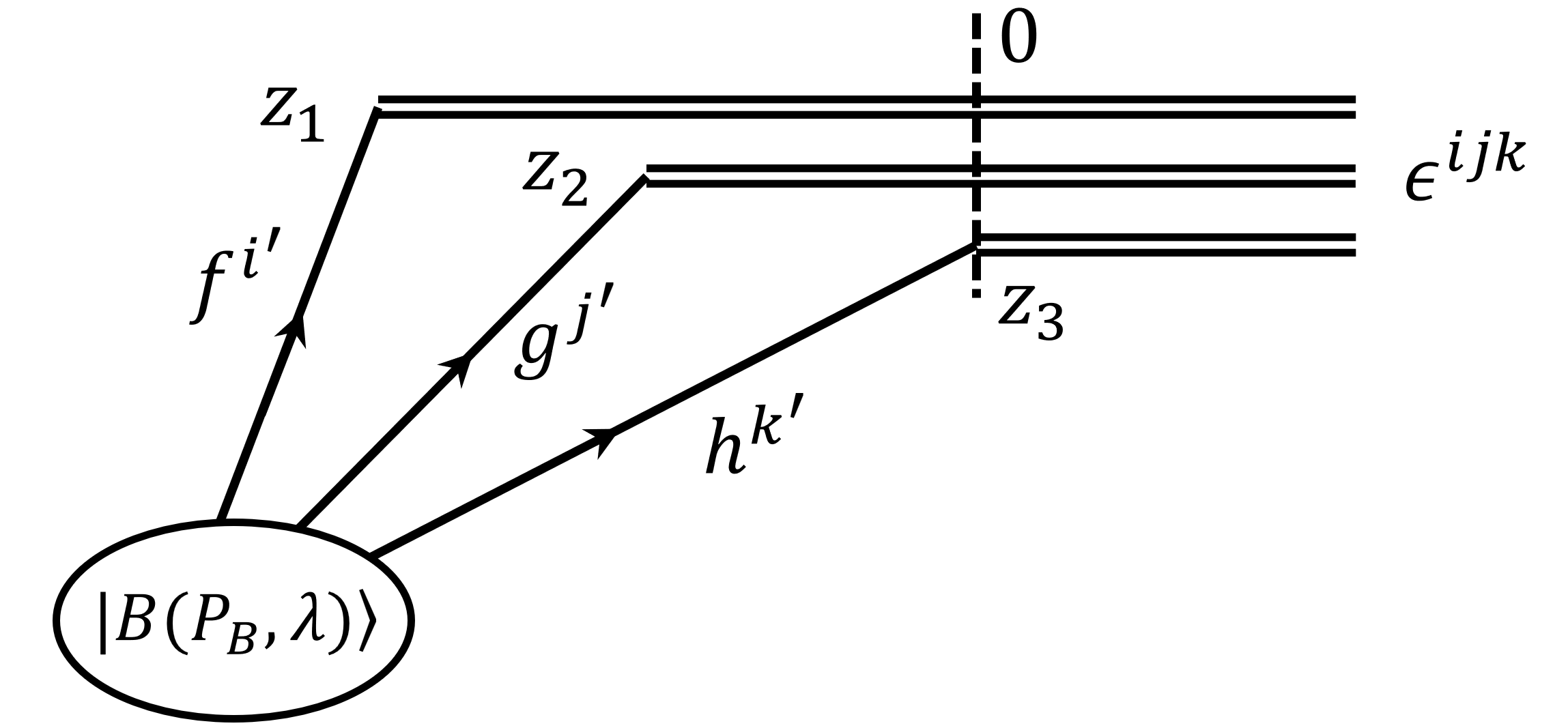}
    }
\vspace{0.0cm} 
\subfigure[LCDAs in momentum space.]{
    \centering
    \includegraphics[scale=0.09]{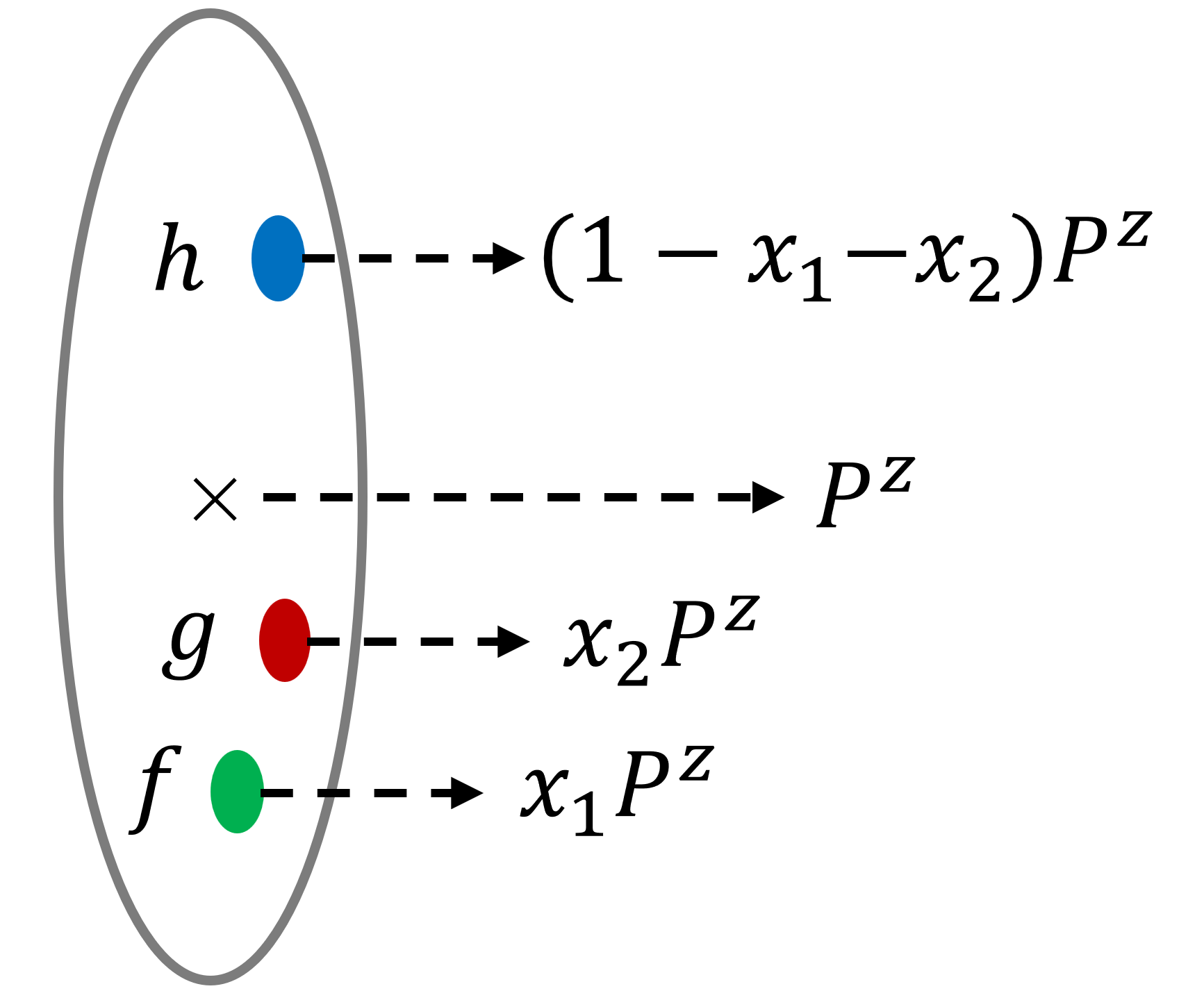}
    }
\caption{The structure of the baryon LCDAs~\cite{LatticePartonCollaborationLPC:2025vhd}.}
\label{fig:structure}
\end{figure*}

Baryon LCDAs are defined through vacuum-to-baryon matrix elements of nonlocal operators with fields separated along the light-cone~\cite{Braun:1999te,Han:2024ucv,Deng:2023csv,Zeng:2025wpd,Shi:2026mjb}:
\begin{equation}
\begin{split}
\epsilon^{ijk} \langle 0 | 
& f_{\alpha}^{i'}(z_1 n) W^{i'i}(z_1 n, z_0 n) \\
\times & g_{\beta}^{j'}(z_2 n) W^{j'j}(z_2 n, z_0 n) \\
\times & h_{\gamma}^{k'}(z_3 n) W^{k'k}(z_3 n, z_0 n) 
| B(P_B) \rangle.
\end{split}
\end{equation}
Here the Wilson lines ensure gauge invariance, and we set $z_3=0$ for convenience, as shown in Fig.~\ref{fig:structure}.

At leading twist, the matrix element can be decomposed into three invariant amplitudes,
\begin{equation}
\begin{split}
\langle 0 | f g h | B(P_B) \rangle
=& \frac{1}{4} f^B \Big[ (\slashed{P}_B C)(\gamma_5 u_B)\, \Phi_V^B \\
& + (\slashed{P}_B \gamma_5 C)(u_B)\, \Phi_A^B \Big] \\
& + \frac{1}{4} f_T^B (i \sigma_{\mu \nu} P_B^{\nu} C)(\gamma^{\mu} \gamma_5 u_B)\, \Phi_T^B .
\end{split}
\end{equation}

The corresponding momentum-space distributions are obtained via Fourier transformation. In this work we concentrate on the A-type LCDA of the $\Lambda$ baryon, which admits a nonzero local limit.

Within LaMET, one replaces light-cone operators by equal-time correlators at large momentum, defining quasi-distribution amplitudes (quasi-DAs). For the A-type structure, the quasi-DA is given by
\begin{equation}
\begin{split}
& {\widetilde{\Phi}_A}^{B}(z_1, z_2, z_3, P_B^{z}) f^B P_B^{z} u_B \\
=& \left\langle 0 \left| f^{T}(z_1 n_z)(C \gamma_5 \gamma^\nu) g(z_2 n_z) h(z_3 n_z) \right| B(P_B) \right\rangle.
\end{split}
\end{equation}

\section{Lattice simulation}\label{sec:setup}

\subsection{Lattice setup}

The calculation is carried out on $N_f=2+1$ ensembles with stout-smeared clover fermions and a Symanzik-improved gauge action generated by the CLQCD collaboration~\cite{CLQCD:2023sdb}. Three lattice spacings are used to eliminate UV divergences and facilitate continuum extrapolation. Information of ensembles are summarized in Table~\ref{tab:ensembles}.
\begin{table}[width=.9\linewidth,cols=4,pos=h]
\caption{Ensembles and lattice setup used in simulation.}\label{tab:ensembles}
\begin{tabular*}{\tblwidth}{@{} LLLL@{} }
\toprule
Ensembles & $a$ (fm) & $N_{\rm cfg} \times N_{\rm src}$ & $P^z$ (GeV)  \\
\midrule
C24P29 & 0.1052 & $864 \times 4$ & 0, 0.49, 1.96 \\
F32P30 & 0.0775 & $777 \times 4$ & 0, 0.50, 2.00 \\
H48P32 & 0.0520 & $550 \times 6$ & 0, 0.50, 1.98 \\
\bottomrule
\end{tabular*}
\end{table}

\subsection{Operators and matrix-elements}

Matrix elements are extracted from two-point correlation functions corresponding to quasi-DAs,
\begin{equation}
\begin{split}
&C_2(t,\vec{P}, z_1,z_2) \\
=&\int d^3xe^{-i\vec{P}\cdot \vec{x}}\langle O_{\rm snk}(\vec{x},t;z_1,z_2)_{\gamma} {\bar O}_{\rm src}(0,0;0,0)_{\gamma'}T^{\gamma'\gamma} \rangle.
\label{eq:2pt_definition}
\end{split}
\end{equation}
The sink operator is chosen to match the quasi-DA structure,
\begin{equation}
\begin{split}
&O^{A}_{\rm snk}(\vec{x},t;z_1,z_2)_\gamma \\
=&\epsilon_{ijk} f_{\alpha}^{i}(\vec{x}+z_1n_z,t) (C \gamma^t)_{\alpha\beta} g_\beta^{j}(\vec{x}+z_2n_z,t) h^{k'}_{\gamma}(\vec{x},t),
\end{split}
\end{equation}
while the source operator is optimized to enhance overlap with the boosted ground state~\cite{Zhang:2025hyo}
\begin{align} \label{eq:source_boost_p_mod}
\begin{split}
O^\Lambda_{\rm src} =& \frac{1}{\sqrt{6}} \Big[ 2(u^TC\gamma_5\gamma^t d)s\\
&+ (u^TC\gamma_5\gamma^t s)d + (s^TC\gamma_5\gamma^t d)u \Big].
\end{split}
\end{align}
The projector $ T = \gamma^t+\gamma^z$ is used to isolate the leading-twist contribution.

To improve signal quality, momentum smeared point source~\cite{Bali:2016lva} and single-step HYP smearing~\cite{Hasenfratz:2001hp,DeGrand:2002vu} of gauge links are applied. The parameterized form of the 2-point function is
\begin{align}
\begin{split}
&\ C^{\rm norm}_2(t,P^z;z_1,z_2) = \frac{C_2(t,P^z;z_1,z_2)}{C_2(t,P^z;0,0)}\\
=&\ \tilde\Phi(z_1,z_2,P^z)\left(1 + A e^{-\Delta E t}\right),
\label{eq:two_state_fit}
\end{split}
\end{align}
the desired matrix elements $\tilde\Phi(z_1,z_2,P^z)$ are obtained from normalized correlators, with excited-state contamination controlled using two-state fits combined with model averaging~\cite{Jay:2020jkz}.

\section{Hybrid renormalization}\label{sec:frame_hy}

\begin{figure*}
	\centering
	\subfigure[$\Lambda$ bare quasi-DA, $z_1 =0$]{
        \centering
        \includegraphics[scale=0.35]{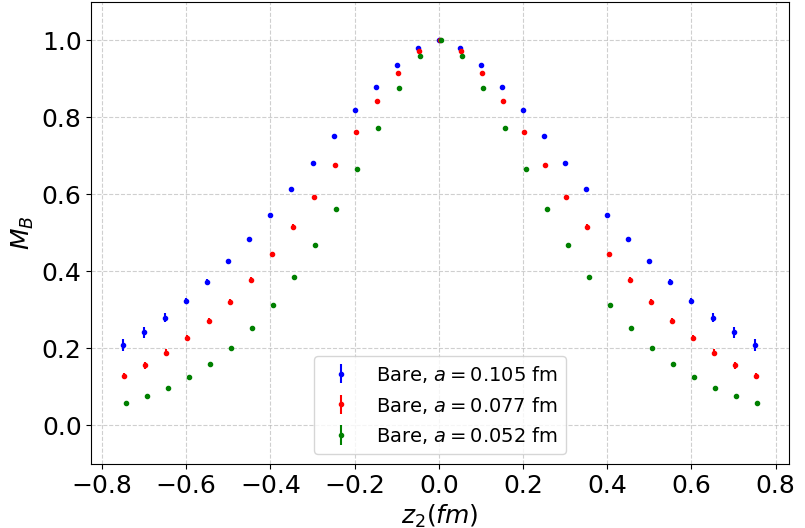}
        }
    \vspace{0.0cm} 
    \subfigure[$1/a$ dependence in logarithmic scale]{
        \centering
        \includegraphics[scale=0.30]{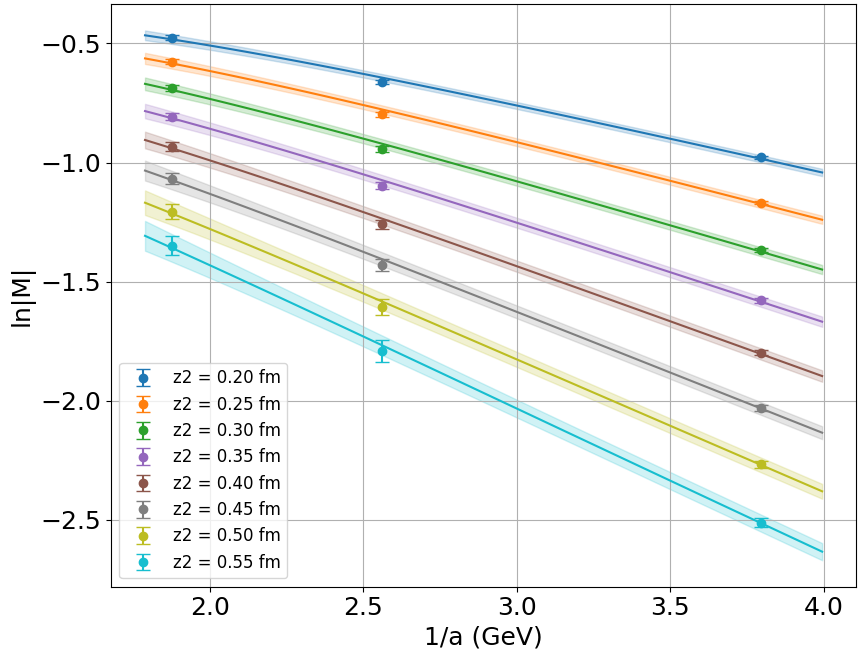}
        }
	\caption{Bare 0-momentum quasi-DAs for $\Lambda$ from 3 different lattice spacings~\cite{LatticePartonCollaborationLPC:2025vhd}.}
	\label{fig:linear_div}
\end{figure*}

Matrix elements computed on the lattice are affected by discretization effects and ultraviolet (UV) divergences. In the left panel of Fig.~\ref{fig:linear_div} we present the bare quasi-DA of the $\Lambda$ baryon obtained at different lattice spacings. The discrepancies among these results are significant and far exceed the expected discretization effects. This behavior becomes more transparent in the logarithmic plot shown in the right panel, where a clear linear dependence is observed.

These features indicate the presence of strong linear divergences that depend explicitly on the lattice spacing, which must be removed in order to achieve reliable matching to the $\overline{\rm MS}$ scheme, and a proper renormalization procedure is therefore essential.

\subsection{Review of renormalization schemes}

The ratio scheme is defined by dividing the large momentum matrix by its zero-momentum counterpart:
\begin{equation}
\hat M_{\rm Ratio}(z_1,z_2;P^z,a) = \frac{\hat M(z_1,z_2;P^z,a)}{\hat M(z_1,z_2;0,a)}.
\end{equation}
As an operator product expansion (OPE)-based approach, the ratio scheme is reliable only in the short-distance region.

To describe the long-distance behavior, the self-renormalization scheme~\cite{LatticePartonLPC:2021gpi} parametrizes the divergences and discretization effects of the bare matrix element as
\begin{equation}
\begin{split}
&\hat M(z_1, z_2, P^z, a) \\
=& \exp \bigg[ \frac{k}{a \ln (a \Lambda_{\rm QCD})} \tilde{z} + g(z_1,z_2,P^z) + f(z_1, z_2) a^2 \\
& + \frac{\gamma_0}{b_0} \ln \frac{\ln (1 /a \Lambda_{\rm QCD})}{\ln (\mu / \Lambda_{\overline{\rm MS}})} 
+ \ln \bigg(1+\frac{d}{\ln (a \Lambda_{\rm QCD})}\bigg) \bigg],
\end{split}
\end{equation}
where $k$ characterizes the linear divergence and $f(z_1,z_2)a^2$ accounts for discretization effects. The function $g(z_1,z_2,P^z)$ encodes nonperturbative physics as well as the mass renormalization contribution~\cite{Ji:1995tm,Zhang:2023bxs}:
\begin{equation}
g(z_1,z_2,P^z) = \ln \hat M_{\overline{\rm MS}}(z_1, z_2, P^z, \mu) + m_0 \tilde{z}.
\end{equation}
At short distances, $\hat M_{\overline{\rm MS}}$ should match the perturbative result $\hat M_p(z_1,z_2,0,\mu)$~\cite{Han:2023xbl}. All parameters in the above expression can be determined through a two-step fit combining lattice data and perturbative input, ultimately yielding the renormalized matrix element in the $\overline{\rm MS}$ scheme.

\subsection{Implementation of hybrid renormalization}
\label{subsec:Hybrid renormalization}

The ratio scheme effectively suppresses ultraviolet divergences at short distances but introduces infrared contamination at large separations. In contrast, the self-renormalization approach removes UV divergences nonperturbatively, yet becomes unreliable in the short-distance region.

\begin{figure}
    \centering
    \includegraphics[width=0.3\textwidth]{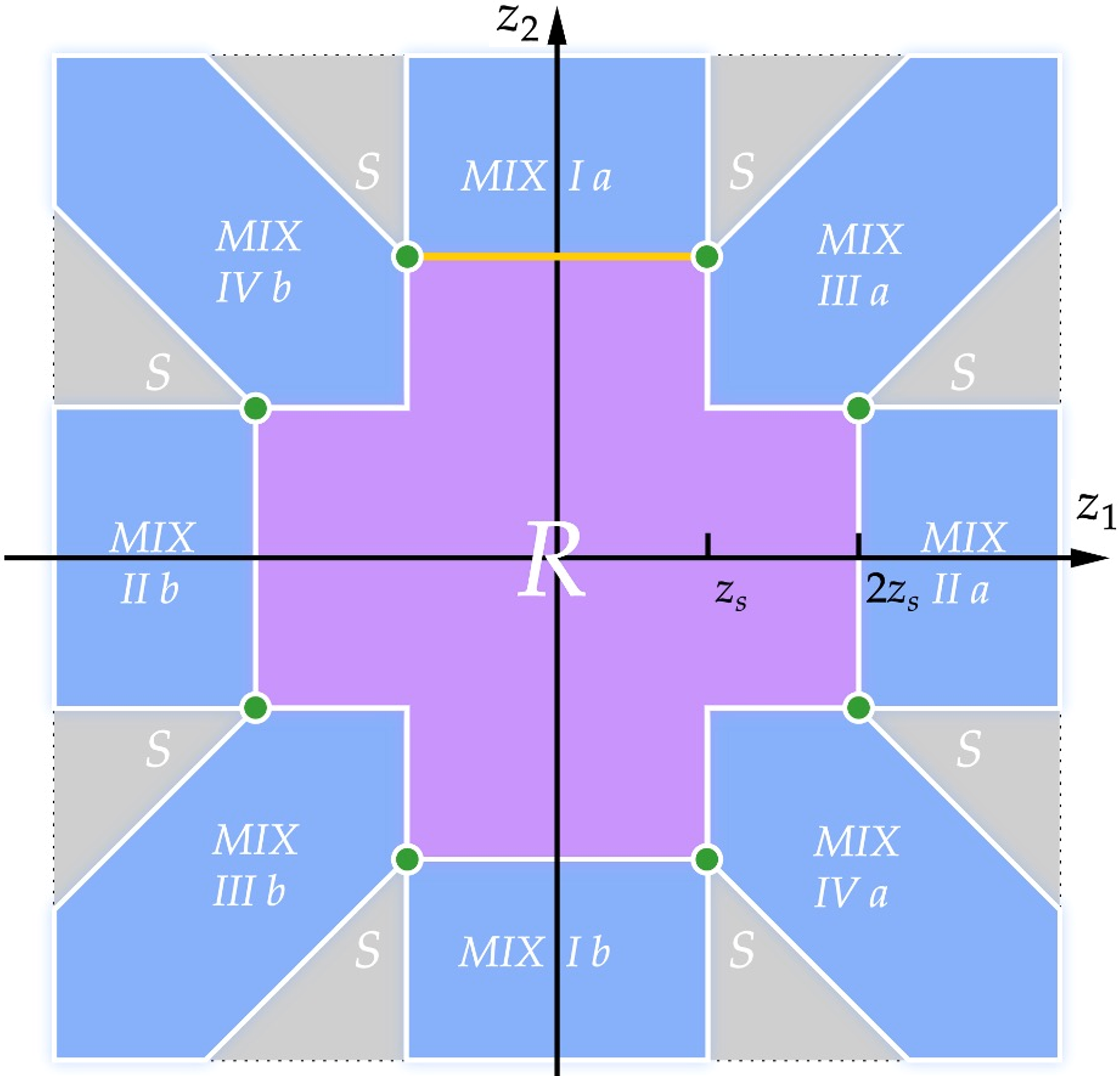}
    \caption{Range division for hybrid renormalization~\cite{LatticePartonCollaborationLPC:2025vhd}.}
    \label{fig:Renorm}
\end{figure}

To combine their advantages, we adopt a hybrid renormalization strategy~\cite{Han:2024ucv,Han:2023xbl,Ji:2020brr,Zhang:2026epr,Zhang:2025npd}, in which different schemes are applied in different regions of the $(z_1,z_2)$ plane. For baryon quasi-DAs, this construction is nontrivial due to the presence of three independent length scales. We therefore partition the coordinate space into short-, long-, and mixed-distance regions, separated by an intermediate scale $z_s \sim 0.2$ fm, as illustrated in Fig.~\ref{fig:Renorm}.

\begin{figure*}
	\centering
	\subfigure[\ Bare result]{
        \centering
        \includegraphics[scale=0.35]{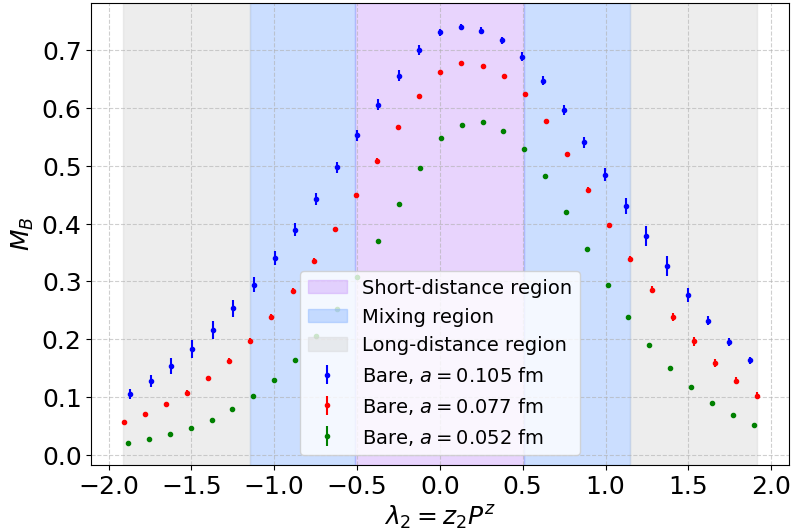}
        }
    \vspace{0.0cm} 
    \subfigure[\ Hybrid scheme result]{
        \centering
        \includegraphics[scale=0.35]{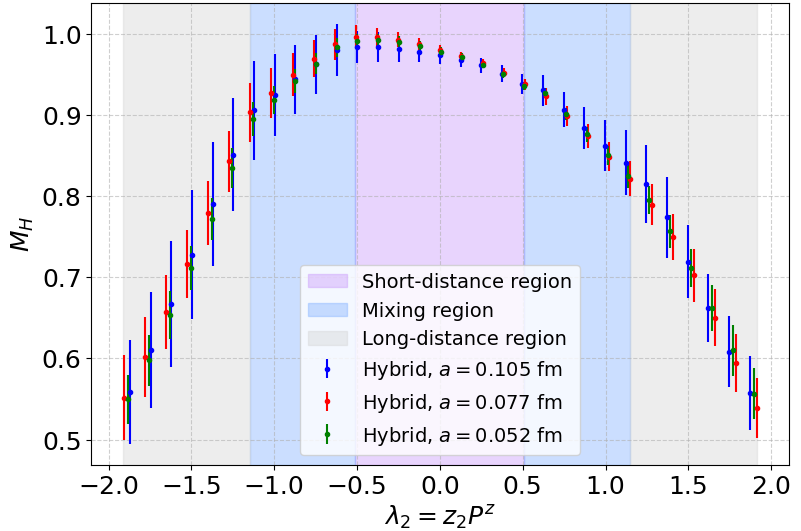}
        }
    \vspace{0.0cm} 
    \subfigure[\ Ratio scheme result]{
        \centering
        \includegraphics[scale=0.35]{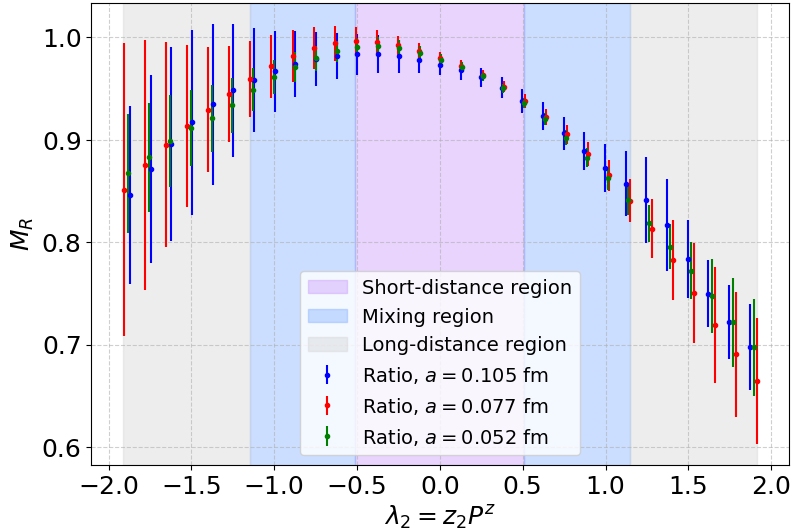}
        }
    \vspace{0.0cm} 
    \subfigure[\ Self-renormalization result]{
        \centering
        \includegraphics[scale=0.35]{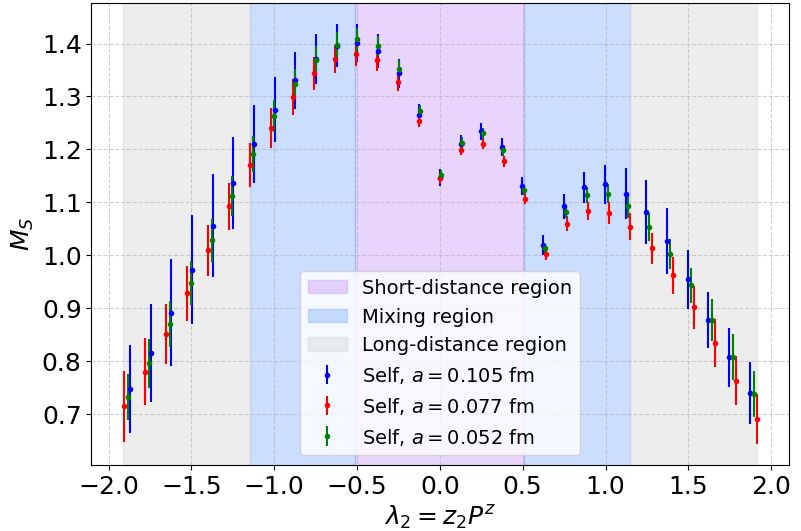}
        }
	\caption{Bare, hybrid, ratio \& self renormalization scheme results of $\Lambda$ quasi-DAs at $P^z=0.5$ GeV~\cite{LatticePartonCollaborationLPC:2025vhd}.}
	\label{fig:renorm_res}
\end{figure*}

In Fig.~\ref{fig:renorm_res} we show the bare and renormalized matrix elements with different schemes. The ratio scheme yields smooth and continuous results, but suffers from extra IR contamination at large distances. The self-renormalization scheme effectively removes UV divergences, yet develops singular behavior in the short-distance region. 

In contrast, the resulting hybrid scheme simultaneously:
\begin{itemize}
\item eliminates linear divergences,
\item avoids short-distance singularities,
\item preserves smooth behavior across all regions.
\end{itemize}

Numerical results demonstrate that this approach provides stable and consistent renormalized matrix elements, improving upon both the ratio and self-renormalization schemes.

\section{Summary}\label{sec:summary}

We reviewed our recent progress in the lattice determination of baryon LCDAs within the LaMET framework. In particular, a hybrid renormalization procedure has been implemented for $\Lambda$ quasi-distribution amplitudes, enabling the removal of linear divergences and a controlled matching to the $\overline{\rm MS}$ scheme.

These developments provide a robust basis for precision studies of baryon structure from first principles. Results for baryon LCDAs at the continuum limit and at physical pion mass are currently being finalized and will be reported in the near future.


\bibliographystyle{model1-num-names}

\bibliography{BaryonLCDAs}


\end{document}